\documentclass[aps,prb,twocolumn,superscriptaddress,showpacs]{revtex4}
\usepackage{graphicx}
\usepackage{amsmath}

\usepackage{color}

\begin{document}

\title{Influence of the Nd $4f$ states on the magnetic behavior and the electric field gradient of the oxypnictides superconductors NdFeAsO$_{1-x}$F$_x$} 

\author{P. Jegli\v{c}} 
\affiliation{Institute ``Jozef Stefan'', Jamova 39, 1000 Ljubljana, Slovenia}

\author{J.-W. G. Bos} 
\email{j.w.g.bos@ed.ac.uk} 
\affiliation{School of Chemistry, University of Edinburgh, West Mains Road, EH9 3JJ, United Kingdom}  

\author{A. Zorko} 
\affiliation{Institute ``Jozef Stefan'', Jamova 39, 1000 Ljubljana, Slovenia}

\author{M. Brunelli} 
\affiliation{European Synchrotron Radiation Facility, BP 220, 38043 Grenoble Cedex, France}

\author{K. Koch}
\affiliation{Max-Planck Institute for Chemical Physics of Solids, N\"othnitzer Str. 40, 01187 Dresden, Germany}

\author{H. Rosner} 
\affiliation{Max-Planck Institute for Chemical Physics of Solids, N\"othnitzer Str. 40, 01187 Dresden, Germany}

\author{S. Margadonna} 
\email{serena.margadonna@ed.ac.uk}
\affiliation{School of Chemistry, University of Edinburgh, West Mains Road, EH9 3JJ, United Kingdom}  

\author{D. Ar\v{c}on}
\email{denis.arcon@ijs.si} 
\affiliation{Institute ``Jozef Stefan'', Jamova 39, 1000 Ljubljana, Slovenia}  
\affiliation{Faculty of Mathematics and Physics, University of Ljubljana, Jadranska 19, 1000 Ljubljana, Slovenia}

 \begin{abstract}
The structural, electronic, and magnetic properties of the superconducting NdFeAsO$_{1-x}$F$_{x}$ phases ($T_C=43\, {\rm K}$ for $x=0.15$) have been investigated experimentally by high-resolution synchrotron x-ray powder diffraction, magnetization and $^{75}$As NMR measurements. Density-functional calculations were performed to calculate and analyze the electric field gradient and the density of states. Compared to LaFeAsO family, the NdFeAsO family shows a contraction of the lattice parameters with shorter rare-earth (RE)-As bond distances, an increased thickness of the As-Fe$_2$-As layer, and less distorted Fe-As$_4$ tetrahedra. The $^{75}$As quadrupole frequencies are  enhanced with respect to the La analogs. This is due to a more prolate As $4p$ electron distribution mainly caused by the reduced lattice parameters and not by the presence of Nd $4f$ electrons. A non-negligible hyperfine coupling between the $^{75}$As nuclei and the Nd $4f$ states indicates a weak coupling between the REO and FeAs layer and possibly opens the channel for a Ruderman-Kittel-Kasuya-Yosida (RKKY)-type interaction between localized Nd $4f$ moments mediated by itinerant Fe $3d$ and/or Nd $5d$ states.
\end{abstract}


\maketitle
\section{Introduction}

The family of rare-earth (RE) quaternary oxypnictides with the general formula REFeAsO (RE = La, Ce, Pr, Nd, Sm, Gd, Tb, Dy) (Ref.~\onlinecite{ena}) is currently generating much excitement in the condensed matter community. When electron doped (e.g., partial replacement of O$^{2-}$ with F$^{-}$ or oxygen deficiency), these materials are reported to display superconductivity with a superconducting transition temperature, $T_C$, as high as 55~K.\cite{dva, tri, stiri, pet, sest, sedem, osem, devet, deset} In electron-doped LaFeAsO the superconductivity ($T_C\sim 26\, {\rm K}$) appears in the close vicinity to a spin-density wave ground state,\cite{dvanajst, osem, stirinajst} implying that antiferromagnetic (AF) correlations may be important in the promotion of the superconducting ground state. La$^{3+}$ has no unpaired electrons, and the magnetic response can be attributed solely to the Fe $3d$ electronic states that dominate the region around the Fermi level. The relation between AF fluctuations and superconductivity has been proposed theoretically \cite{sestnajstA, sedemnajstA} and addressed experimentally by$^{19}$F, $^{75}$As, and $^{139}$La NMR \cite{sestnajst, LaNuQ, osemnajst}. Upon F doping, a pseudo-gap behavior has been found in the temperature dependence of the $^{19}$F (Ref. \onlinecite{sestnajst}) and $^{75}$As (Ref.~\onlinecite{LaNuQ}) spin-lattice relaxation rates.

Substitution of La$^{3+}$ with other rare-earth elements such as Nd$^{3+}$ (free-ion magnetic moment $\mu\sim 3.6\, \mu_B$) introduces additional spin degrees of freedom in the REO layer due to the unpaired RE $4f$ electrons. Interestingly, the NdFeAsO$_{1-x}$F$_x$ family displays superconductivity with enhanced $T_C$'s of $\sim 50\, {\rm K}$ compared to the La-based materials.\cite{dva, sedem} This raises important questions of if and how the rare-earth magnetic moment interacts with the conducting FeAs layer and whether these 4$f$ moments or lattice effects (chemical pressure) have a stronger influence on the electronic structure and the related magnetic and superconducting properties in the REFeAsO family.  

To address these questions, we have studied the structural, magnetic, and superconducting properties of the NdFeAsO$_{1-x}$F$_x$ family by means of high-resolution synchrotron x-ray powder diffraction, magnetization, and $^{75}$As NMR measurements. Our experimental results are supported by electronic structure calculations.

\section{methods}

Polycrystalline NdFeAsO and NdFeAsO$_{0.85}$F$_{0.15}$ samples were prepared by a standard two-step solid-state reaction method. The magnetic susceptibility was measured using a Quantum Design magnetic property measurement system. Synchrotron powder diffraction data sets, collected on the ID31 high-resolution diffractometer at the European Synchrotron Radiation Facility in Grenoble, France, were  binned with a $0.002^\circ$ stepsize ($\lambda= 0.309952$~\AA). The GSAS suite of programs and EXPGUI graphical user interface were used for Rietveld fitting.\cite{GS1,GS2} $^{75}$As ($I=3/2$) NMR frequency-swept spectra were measured in 8.9~T magnetic field with a selective solid echo pulse sequence (pulse length of $\pi /2 = 22~\mu{\rm s}$ and delay of $\tau = 50~\mu {\rm s}$). The reference frequency of $\nu (^{75}{\rm As}) = 65.096~{\rm MHz}$ was determined from a NaAsF$_6$ standard.
 
The band-structure calculations were performed using the full-potential local-orbital minimum basis code FPLO (version 5.00-19) (Ref.~\onlinecite{FPLO}) within the local density approximation(LDA). In the scalar relativistic calculations the exchange and correlation potential of Perdew and Wang \cite{PW} was employed. To treat the Nd $4f$ state adequately LSDA+$U$ in the atomic limit was employed with $U^{\rm Nd}=[7,9]~$eV and $J=0.7$~eV. As basis set Nd (4p4d4f5s5p/6s6p5d+7s7p), Fe (3s3p/4s4p3d+5s5p), As (3p/4s4p3d+5s5p) and O (2s2p3d+3s3p) were chosen for semicore/valence+polarization states. The high lying states improve the basis which is especially important for the calculation of the electric field gradient (EFG) tensor with components $V_{ij}=\partial V/\partial x_i\partial x_j$. The lower lying states were treated fully relativistic as core states. A well converged $k$ mesh of 252 $k$ points was used in the irreducible part of the Brillouin zone. In order to investigate the influence of F substitution on the O site, the virtual crystal approximation (VCA) was applied.

\section{results}

\begin{figure} [htbp]
\includegraphics* [width=\columnwidth] {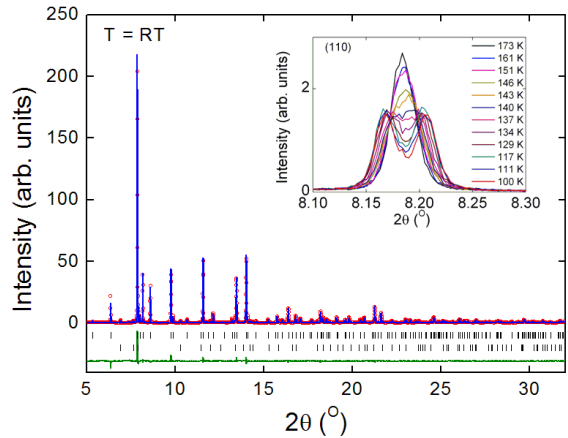}
\caption{(Color online) Measured ($\bigcirc$) and calculated (--) synchrotron x-ray powder diffraction ($\lambda= 0.309952 $ \AA, room temperature) profiles for NdFeAsO. The lower solid line shows the difference profile, and the tick marks show the reflection positions (top: NdFeAsO, bottom: Nd$_2$O$_3$ impurity). The inset illustrates the $T$-$O$ structural phase transition.}
\label{Figure1}
\end{figure}

High-resolution synchrotron x-ray powder diffraction profiles of NdFeAsO at room temperature confirmed that its structure is tetragonal---space group $P4/nmm$---in analogy to other REFeAsO systems.  Rietveld analysis of the synchrotron X-ray powder diffraction data (Fig.~\ref{Figure1}) resulted in lattice parameters {$a = 3.96629(1)$~\AA}\, and {$c = 8.59886(6)$~\AA}\, with Fe-As and Fe-Fe distances of 2.4007(4)~\AA\, and {2.80459(1)~\AA}, respectively and an As-Fe-As bond angle of {$111.39(3)^\circ$} [$\chi^2=2.4$, $wRp = 13.2\%$, $Rp = 9.2\%$, Nd: {$z = 0.13887(6)$}, and As: {$z = 0.65735(9)$}]. \cite{therm} Compared to LaFeAsO,\cite{dvanajst} the Nd$^{3+}$ containing oxypnictide shows a contraction of the lattice constants with shorter RE-As bond distances, an increased thickness of the As-Fe$_2$-As layer and less distorted Fe-As$_4$ tetrahedra. Temperature-dependent synchrotron x-ray powder diffraction measurements collected between room temperature and 100 K revealed the presence of a structural phase transition to an orthorhombic ($O$) superstructure ($a_O\sim b_O=\sqrt{2}a_T$ and $c_O=c_T$; space group $C mma$) in analogy with the results reported in Ref.~\onlinecite{Bianconi}. All the Bragg reflections containing the tetragonal Miller indexes  $h\neq 0$ and $k\neq 0$ start broadening at 160 K and finally split only at 140 K (inset to Fig.~\ref{Figure1}).  
 
The F-doped compound, with nominal composition NdFeAsO$_{0.85}$F$_{0.15}$, also adopts tetragonal symmetry but with reduced room temperature lattice parameters [{$a$ = 3.96251(1)~\AA}\, and {$c$ = 8.57785(6)~\AA})] and Fe-As and Fe-Fe distances of 2.4025(6)~\AA\, and {2.80192(1)~\AA}, respectively ($\chi^2=2.4$, $wRp = 12.84\%$, $Rp = 10.31\%$, Nd: {$z = 0.14211(6)$}, and As:  $z = 0.6544(1)$). We notice that despite the smaller ionic radius of F$^-$ compared to O$^{2-}$, the RE-O distance increases, resulting in an increased thickness of the Nd-O$_2$-Nd layer and more regular Nd$_4$-O tetrahedra.

\begin{figure} [htbp]
\includegraphics* [width=0.7\columnwidth] {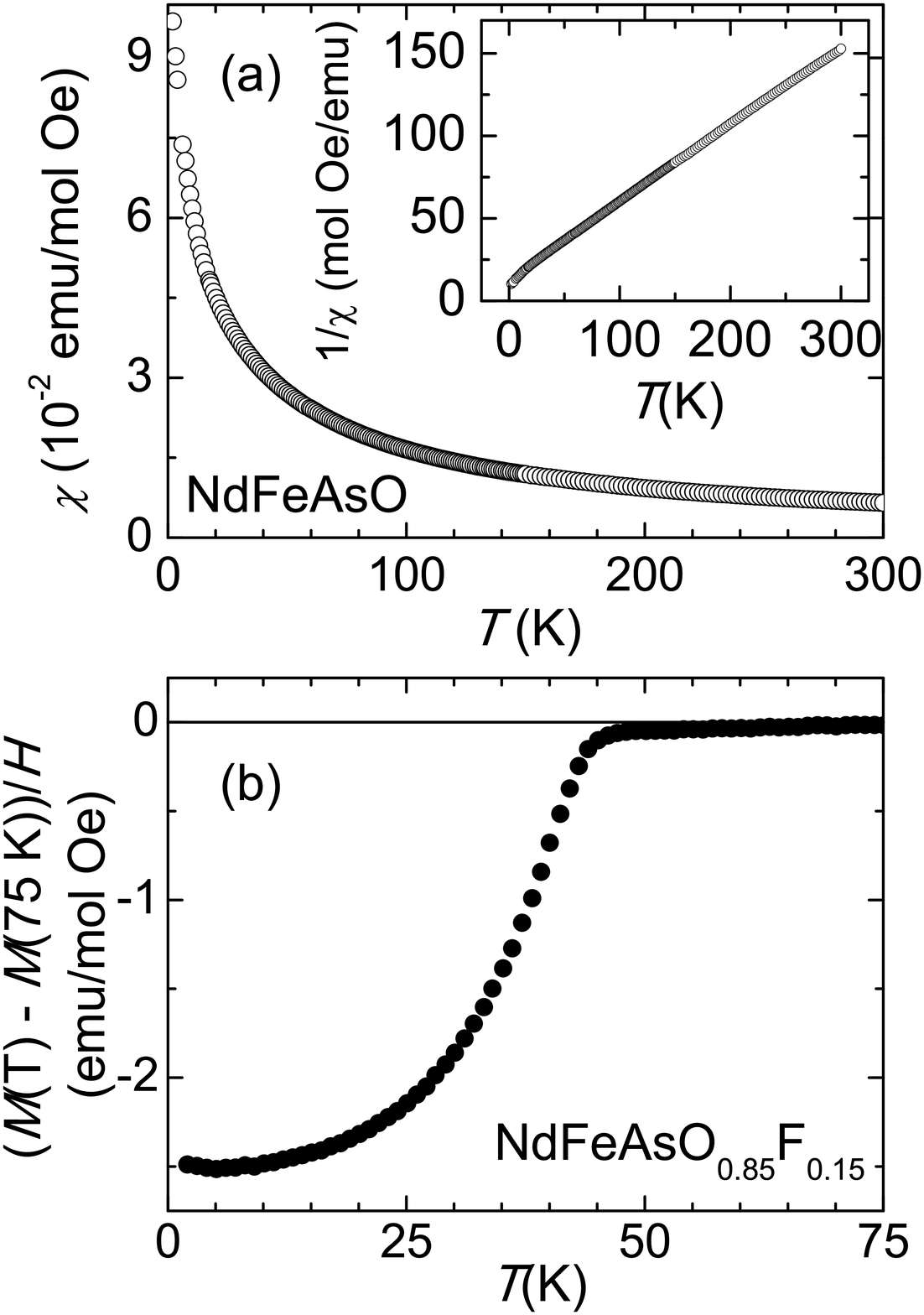}
\caption{(a) Temperature dependence of the ZFC susceptibility measured in 10 kOe applied field for NdFeAsO powder. Inset: temperature dependence of the inverse ZFC susceptibility for NdFeAsO powder.(b) Temperature dependence of the ZFC magnetization for NdFeAsO$_{0.85}$F$_{0.15}$ with the superconducting transition temperature of 43 K and a shielding fraction of 40.6\%. The magnetic field was set to 20 Oe.}
\label{Figure2new}
\end{figure}

The zero-field-cooled (ZFC) molar magnetic susceptibility, $\chi$, of NdFeAsO measured in a 10 kOe applied field shows a substantial Curie-Weiss-type contribution [Fig.~\ref{Figure2new}a] that completely dominates $\chi$ at low temperatures. Analysis of the susceptibility data gives the Curie constant $C=1.43(2)~{\rm emuK/mol}$ and the Curie-Weiss temperature $\theta=-14(1)~{\rm K}$ [inset to Fig.~\ref{Figure2new}a)]. The Curie constant corresponds to an effective magnetic moment $\mu_{eff}=3.31(4)~\mu_B$ in agreement with the expected value for localized Nd $4f$ moments, while the negative Curie-Weiss temperature implies AF interactions between the Nd $4f$ moments. We also note here that the concentration of Nd$_2$O$_3$ impurities is so small that the values of $C$ and $\theta$ do not change after molar susceptibility is corrected for the impurity contribution. The temperature dependence of the zero-field-cooled magnetization measured on NdFeAsO$_{0.85}$F$_{0.15}$ [Fig.~\ref{Figure2new}b] shows a clear onset of bulk superconductivity at $T_C = 43(1)~{\rm K}$ with shielding fraction of 40\%.

\begin{figure} [htbp]
\includegraphics* [width=\columnwidth] {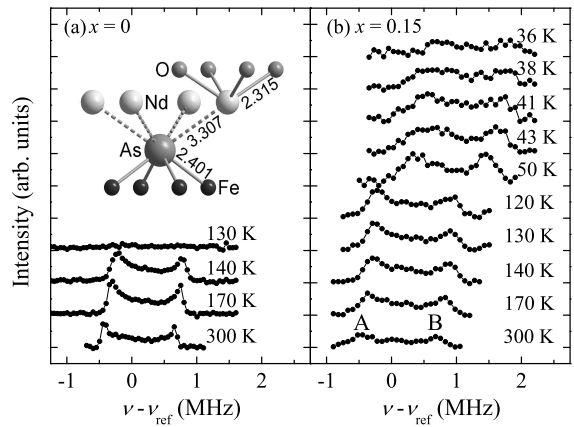}
\caption{Representative $^{75}$As NMR spectra measured in (a) the parent NdFeAsO and (b) the F-doped NdFeAsO$_{0.85}$F$_{0.15}$ powder samples. Inset: structure around the As ion showing its room-temperature distances to nearest-neighboring Fe (black spheres) and Nd ions (white spheres).}
\label{Figure2}
\end{figure}

Representative powder $^{75}$As NMR spectra of NdFeAsO and NdFeAsO$_{0.85}$F$_{0.15}$ are shown in Fig.~\ref{Figure2}. The room-temperature spectra for both samples display a typical quadrupole powder line shape. For undoped NdFeAsO the shape of the central $-1/2 \leftrightarrow 1/2$ transition and the sharp singularities for the satellite transitions strongly indicate that the asymmetry parameter, defined by the components of the EFG tensor as $\eta =\left( V_{xx}-V_{yy}\right) /V_{zz}$, is $\eta=0$. $V_{zz}$ defines the quadrupole frequency 
\begin{eqnarray}
\label{nuQ}
\nu_Q={3\over 2}{eV_{zz}Q\over I(2I-1)},
\end{eqnarray}
with $Q$ as the quadrupole moment of $^{75}$As. We recall that the $^{75}$As nuclei have four nearest-neighbor Fe atoms (inset to Fig.~\ref{Figure2}) and in the tetragonal phase reside on the 2c $(1/4, 1/4 , z)$ position, which is fourfold axially symmetric. $\eta=0$ is therefore fully consistent with the As site symmetry.  The $^{75}$As powder NMR line shape simulation (inset to Fig.~\ref{Figure3}) then gives $\nu_Q = 11.8(1)~{\rm MHz}$.  We note that $\nu_Q$ is significantly enhanced compared to $\nu_Q=8.7(4)$~MHz (Ref.~\onlinecite{HM}) for LaFeAsO and it is even larger than $\nu_Q = 11.00(5)~{\rm MHz}$ reported for LaFeAsO$_{0.9}$F$_{0.1}$.\cite{LaNuQ} This is presumably a result of slightly different charge distributions around the As site reflecting the small structural variations in the As local environment as it will be discussed below. The $^{75}$As NMR spectra of NdFeAsO$_{0.85}$F$_{0.15}$ are considerably broader compared to the parent compound, reflecting local disorder introduced by the F doping. This prevents a precise determination of $\eta$. However, the $^{75}$As NMR spectra can be simulated using a Gaussian distribution of $\nu_Q$ with a mean value of $\nu_Q = 12.9(2)~{\rm MHz}$ and a full width at half height of the Gaussian distribution $\Delta\nu_Q=1.8(2)~{\rm MHz}$. 
   
\begin{figure} [htbp]
\includegraphics* [width=\columnwidth] {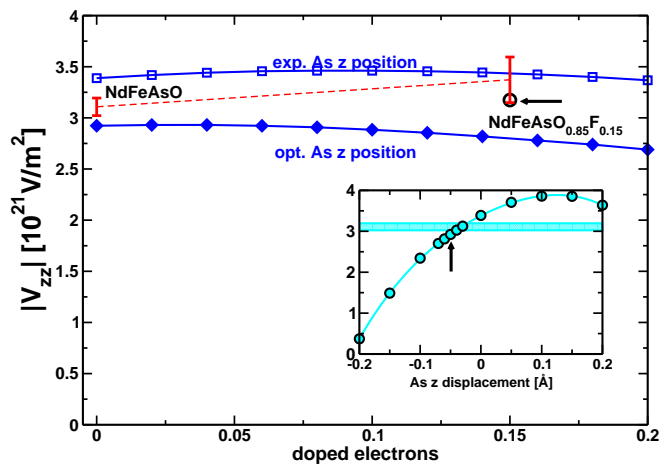}
\caption{(Color online) Calculated $V_{zz}$ obtained from the virtual crystal approximation using the experimental As {$z = 0.65735(9)$} position (empty squares), the optimized As $z=0.6515$ (filled diamonds) position, and the NdFeAsO$_{0.85}$F$_{0.15}$ structure at room temperature (black circle marked by an arrow). The measured EFGs for the pure and the 15\% F-doped compound are shown by error bars. Inset: dependence of $^{75}$As $V_{zz}$ on the As $z$ position. The optimized As $z$ position is marked by an arrow. The experimental $V_{zz}$ is represented by the shaded bar.}
\label{FigureEFG}
\end{figure}

For the undoped compound, the measured EFG is obtained by applying the $^{75}$As quadrupole moment \cite{pyykko} $Q=(0.314\pm0.006)$~b in Eq.~(\ref{nuQ}) and yields $|V_{zz}^{exp}| = (3.11\pm0.09)\cdot 10^{21}$~V/m$^2$. Using the experimental room temperature lattice parameters and atomic positions and $U^{\rm Nd}=8$~eV we obtain a good agreement for the calculated EFG: $V_{zz}^{calc}=-3.39 \cdot 10^{21}$~V/m$^2$. Varying $U^{\rm Nd}$ by $\pm1$~eV (within the physically reasonable range) changes the EFG only by $\mp 0.01\cdot 10^{21}$~V/m$^2$, which is well below the experimental error bars. Like for the Fe magnetic moment, which in the calculations shows a strong dependence on the As $z$ position, \cite{CK, KK} we observe also for the EFG a strong As $z$ dependence (inset to Fig.~\ref{FigureEFG}). The calculated EFG ($V_{zz}^{calc}=-2.92\cdot 10^{21}$~V/m$^2$) is even closer to the measured one when As is shifted along the negative $z$ direction to  $z=0.6515$. Here, the energy has a minimum (in inset to Fig.~\ref{FigureEFG} marked by an arrow) and the structure has a shorter Fe-As distance of $2.3729$~{\AA}.

Analogous to the experiments, where the quadrupole frequency for $^{75}$As is larger for the Nd than for the La compound, the magnitude of the calculated $| V_{zz} |$ is larger for the Nd compound $V_{zz}=-3.39\cdot 10^{21}$~V/m$^2$ than for the La compound \cite{LaHajo} $V_{zz}=-3.21\cdot 10^{21}$~V/m$^2$ but less drastically. Analysis of the calculated EFG (Ref.~\onlinecite{STOBTO}) shows that the largest contribution to the EFG mainly originates from As $4p$ electrons.  In both compounds the anisotropy count $\Delta_p =\frac{1}{2}(n_x+n_y)-n_z$ for the As $4p$ electrons is negative, which corresponds to a prolate $p$ electron distribution and is in agreement with the negative $V_{zz}$. For the La compound $\Delta_p$ is less negative than for the Nd compound: $\Delta_p=-0.018$ and $\Delta_p=-0.062$, respectively, which means that the $4p$ orbitals are less isotropic for the Nd compound and hence the EFG is enhanced. The observed difference between the La and Nd compounds could arise from either the additional 4$f$ electrons due to screening effects or from the smaller lattice parameters in the Nd compound. To separate the influence of the change in the lattice geometry from the change in the electronic configuration, we compare the real EFGs with fictitious EFGs, which are obtained  by exchanging the lattice parameters for both compounds (see Table~I).

\begin{table}[tbp]
\caption{Calculated $V_{zz}$ for As for both compounds for both lattice parameters. In the calculations we used the following room-temperature lattice parameters: set~1 [LaFeAsO (Ref.~\onlinecite{dvanajst})]: $a$=4.03007(9)~\AA\,, $c$=8.7368(2)~\AA\,, As: $z=0.6507(4)$, and La: $z=0.1418(3)$; set~2 (NdFeAsO, this work): {$a$ = 3.96629(1)~\AA\, and $c$ = 8.59886(6)~\AA\,; As: $z = 0.65735(9)$ and Nd: $z =0.13887(6)$}. LaFeAsO with set~1 and NdFeAsO with set~2 are the real compounds; the other combinations are fictitious.}
\begin{tabular}{ccc}
\hline
\hline
& Set~1 & Set~2
\\
\hline
LaFeAsO  & $-3.21 \cdot 10^{21}$~V/m$^2$ & $-3.60 \cdot 10^{21}$~V/m$^2$
\\
NdFeAsO  & $-2.92 \cdot 10^{21}$~V/m$^2$ & $-3.39 \cdot 10^{21}$~V/m$^2$
\\
\hline
\hline
\end{tabular}
\end{table}

The $|V_{zz}|$ value increases with lattice compression both for the LaFeAsO and NdFeAsO electronic configurations (Table~I from the left to the right). Next, we checked  the role of the RE $4f$ electrons. In this case we find the opposite trend; i.e., for given lattice parameters $|V_{zz}|$ is smaller for the Nd than for the La compound (Table~I from top to bottom). One can therefore conclude that $|V_{zz}|$  increases with lattice contraction and decreases with increasing number of RE $4f$ electrons.

Now, we focus on the effect of electron doping on the EFG in NdFeAsO. The EFGs of the doped compounds were calculated with the virtual crystal approximation. The validity of the VCA was confirmed by supercell calculations.\cite{supercell} First, we consider solely  the effect of electron doping. Therefore, we keep the structural parameters fixed for different levels of doping. In Fig.~\ref{FigureEFG} two such VCA curves are shown. When the experimentally determined As $z = 0.65735(9)$ position (NdFeAsO at room temperature) is used, the calculated and measured EFG values for 15\% doping agree very well. The VCA curve with the optimized As $z=0.6515$ position deviates from the experimental curve as it systematically predicts a smaller $|V_{zz}|$. Next, we investigate the structural change on top of the doping by calculating the EFG within VCA for the room-temperature structural parameters of NdFeAsO$_{0.85}$F$_{0.15}$. We notice that the use of these parameters slightly reduces $|V_{zz}|$ to $V_{zz}=-3.17\cdot 10^{21}$~V/m$^2$. Please note the black circle marked by an arrow in Fig.~\ref{FigureEFG}. We stress at this point, that the effect of electron doping on the EFG is much smaller than the influence of the As $z$ position as can be clearly seen by comparing Fig.~\ref{FigureEFG} with the inset of Fig.~\ref{FigureEFG}. Our calculations predict a small decrease in $|V_{zz}|$ upon electron doping for NdFeAsO$_{0.85}$F$_{0.15}$, although experimentally a slight increase is observed. This is also the case for the La compound where the difference between the slopes of the calculated and the experimental EFGs upon doping \cite{LaHajo} is even more pronounced. Further studies are required to investigate if this is due to the quality of the samples or due to intrinsic changes in the electronic structure.

In undoped NdFeAsO the $^{75}$As NMR line shape starts to change below $T_S\sim 160~{\rm K}$ (Fig.~\ref{Figure2}).  Its evolution can be accounted for by assuming that $\eta$ suddenly starts to increase below $T_S$ and reaches $\eta = 0.10(2)$ at $140~{\rm K}$. The increase in $\eta$ is a strong indication that the $^{75}$As site symmetry has lowered due to the structural phase transition to a low-temperature orthorhombic structure (space group $Cmma$),\cite{Bianconi} similarly as found in other REOFeAs systems. We note that in the orthorhombic phase $^{75}$As is on the 4$g$ $(0, 1/4, z)$ position, allowing $\eta$ to differ from zero. The band-structure calculations show that the influence of the orthorhombic distortion on the EFG is negligible -- much less than the experimental error bars. This is understandable given the small size of the orthorhombic splitting. The anisotropy parameter is more sensitive to the distortion and increases to $\eta=0.04$ (compare to the experimental value of $\eta=0.10(2)$). Below $T_N=140~{\rm K}$ the high-temperature $^{75}$As resonance in NdFeAsO starts to rapidly lose in intensity on the account of an extremely broad low-temperature signal (inset to Fig.~\ref{Figure3}). The dramatic broadening of the $^{75}$As resonance is a manifestation of AF magnetic order of Fe moments occurring at $T_N=140~{\rm K}$, which was previously observed by $\mu$SR (Ref.~\onlinecite{Aczel}) and neutron-diffraction \cite{chen} measurements. In the doped NdFeAsO$_{0.85}$F$_{0.15}$ both the structural and AF transitions are absent at least down to the superconducting transition temperature $T_C=43~{\rm K}$, below which the $^{75}$As resonance rapidly disappears (Fig.~\ref{Figure2}). Lineshape simulations also show that $\nu_Q$ is almost temperature independent and therefore cannot account for the large temperature dependence of the observed $^{75}$As NMR shift.

\section{discussion}

\begin{figure} [htbp]
\includegraphics* [width=\columnwidth] {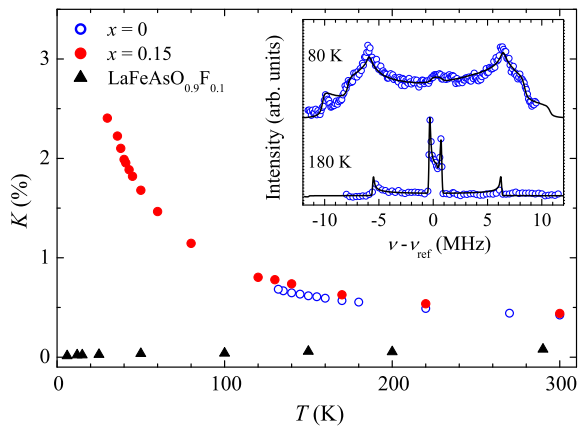}
\caption{(Color online) Temperature dependence of the $^{75}$As NMR shift measured for undoped NdFeAsO (open circles) and F-doped NdFeAsO$_{0.85}$F$_{0.15}$ powder samples (solid circles). For comparison we show the corresponding shift measured in LaFeAsO$_{0.9}$F$_{0.1}$ (solid triangles) (Ref.~\onlinecite{LaNuQ}).  Inset: the $^{75}$As NMR line shape simulations [Eq. 1] for NdFeAsO sample above ($T=180\, {\rm K}$) and below ($T=80\, {\rm K}$) the AF ordering transition $T_N=140~{\rm K}$.} 
\label{Figure3}
\end{figure}

In order to understand the observed $^{75}$As resonance for the undoped NdFeAsO in the AF phase, we assume that this phase is characterized by the spin-density wave vector $\vec Q=(1,0,1)$ and that the ordered Fe magnetic moments are aligned along the $a$ axis, i.e., similarly as in LaFeAsO.\cite{osem} $\mu$SR measurements suggested that the Fe ordered moment in NdFeAsO is similar to that of LaFeAsO, i.e., $\mu_{\rm Fe}\sim 0.3~\mu_B$.\cite{Aczel} The shift of the $^{75}$As resonance in the AF phase is due to the internal field $B_{\rm int}$ aligned along the crystal $c$ axis and is given by the anisotropic (off-diagonal) part of the hyperfine interaction with Fe moments, $a_{xz}$, as $B_{\rm int}=4a_{xz}~\mu_{\rm Fe}$.\cite{Kitagawa} Its isotropic part in the ordered phase is filtered out from the shift due to the highly symmetrical position of $^{75}$As with respect to Fe moments.  The projection of this field adds to the external magnetic field, corrected for an isotropic hyperfine shift caused by Nd moments, $B_0$ and gives rise to the effective magnetic field $B_{\rm eff}=B_0+B_{\rm int}\cos\vartheta$. Here, $\vartheta$ denotes the angle between the external and internal magnetic fields. Taking into account also the first-, $\Delta\nu_Q^{(1)}(\vartheta,\phi)$, and the second-order, $\Delta\nu_Q^{(2)}(\vartheta,\phi)$, quadrupole effects, the $^{75}$As resonance frequency reads as
\begin{equation}
\nu_{m\leftrightarrow  m-1}
=\gamma_{\rm As}B_{\rm eff}+\Delta\nu_Q^{(1)}(\vartheta,\phi)
+\Delta\nu_Q^{(2)}(\vartheta,\phi)\,.
\end{equation}
Here, $\gamma_{\rm As}$ is the $^{75}$As gyromagnetic ratio and $\vartheta$ and $\phi$ define the orientation of $B_{\rm eff}$ with respect to the principal axes of EFG tensor. Our simulation with Eq. (1) results in a good fit of the 80 K spectrum when $a_{xz}={10.5~{\rm kOe}/ \mu_B}$ was used (inset to Fig.~\ref{Figure3}). We stress at this point that the large anisotropic hyperfine fields cannot be explained by dipolar fields of the Fe moments and that they rather provide experimental evidence for the Fe $3d$ and As $4p$ hybridizations. \cite{enajst, Kitagawa}

We now turn to the determination of the isotropic part of the $^{75}$As hyperfine interaction measured by the Knight shift, $K$, in the paramagnetic phase (Fig.~\ref{Figure3}). We stress that the measured shift is by an order of magnitude larger and has a completely different temperature dependence when compared to the one for LaFeAsO$_{0.9}$F$_{0.1}$.\cite{LaNuQ} Temperature dependence of $K$ in NdFeAsO$_{1-x}$F$_{x}$ can be  well described by the Curie-Weiss-type behavior, $K(T) = K_0 + \kappa/(T - \theta)$.  The value of the Curie-Weiss temperature $\theta=-13(2)~{\rm K}$ [$\kappa = 0.99(2)~{\rm K}$] is the same as determined from the magnetic susceptibility. We can therefore assume that the Curie-Weiss part of the shift comes from the Fermi contact hyperfine interaction between the $^{75}$As nuclei and Nd $4f$ moments. The value of $K_0 = 0.096(15)\%$ is close to the total NMR shift reported for LaFeAsO$_{0.9}$F$_{0.1}$.\cite{LaNuQ} We thus associate this part of the $^{75}$As NMR shift to the coupling of $^{75}$As with the itinerant Fe $3d$ electrons in the FeAs layer. Interestingly, for undoped NdFeAsO we find that $K_0$ is slightly reduced, i.e., $K_0 = 0.091(20)\%$.

We proceed with the above analysis by recognizing that $K_0$ is given by the hyperfine coupling to itinerant FeAs electrons and the orbital part, 
\begin{eqnarray}
K_0 = {a_{\rm FeAs}^{\rm iso}\over N_A\mu_B}\chi_{\rm FeAs}+K_{\rm orb}.
\end{eqnarray}
Here $K_{\rm orb}=-0.075\%$ is the orbital shift,\cite{LaNuQ} $N_A$ and $\mu_B$ are the Avogadro number and Bohr magneton, $\chi_{\rm FeAs}$ is the molar susceptibility for the itinerant FeAs electrons, and finally $a_{\rm FeAs}^{\rm iso}$ is the corresponding isotropic hyperfine coupling constant. Since a precise dependence of $\chi_{\rm FeAs}$ is in NdFeAsO$_{1-x}$F$_{x}$ completely masked by the magnetism of the Nd $4f$ moments (Fig.~\ref{Figure2new}), we approximate it with the dependence for LaFeAsO taken from Ref.~\onlinecite{LaNuQ}. Using the value of $K_0$ we derive $a_{\rm FeAs}^{\rm iso}=27.7~{\rm kOe}/\mu_B$ and $a_{\rm FeAs}^{\rm iso}=29~{\rm kOe}/\mu_B$ for NdFeAsO and NdFeAsO$_{0.85}$F$_{0.15}$, respectively.  Although these values have to be taken with certain caution due to approximations used, we notice that they are about 10\%-15\% larger than in LaFeAsO$_{0.9}$F$_{0.1}$.\cite{LaNuQ}

The contribution arising from the coupling to Nd $4f$ moments, $K_{\rm Nd}$, can be expressed as
\begin{eqnarray}
K_{\rm Nd} = {\kappa /(T-\theta)}={a_{\rm Nd}^{\rm iso}\over N_A\mu_B}\chi_{\rm Nd},
\end{eqnarray}
where $\chi_{\rm Nd}=C/(T-\theta)$ is the molar susceptibility of localized Nd $4f$ moments and $a_{\rm Nd}^{\rm iso}$ is the corresponding isotropic hyperfine coupling constant. Using the value of $\kappa$ we calculate $a_{\rm Nd}^{\rm iso}=3.8~{\rm kOe}/\mu_B$.  Although $a_{\rm Nd}^{\rm iso}$ is by an order of magnitude smaller than $a_{\rm FeAs}^{\rm iso}$, it is still surprisingly large as the REFeAsO system has so far been believed to comprise well isolated Fe-As and REO layers.\cite{enajst}  

The coexistence of itinerant Fe $3d$ and localized Nd $4f$ moments in NdFeAsO$_{1-x}$F$_{x}$ is highly intriguing, and one wonders how the magnetic and/or superconducting properties are affected. Our experimental results clearly prove that the $^{75}$As NMR probe interacts both with the Fe $3d$ conducting electrons in the FeAs layer as well as with the localized Nd $4f$ moments in the NdO layer. The sizeable anisotropic hyperfine coupling constant between $^{75}$As and Fe $3d$ moments ($a_{xz}={10.5~{\rm kOe}/ \mu_B}$) in the AF phase has its origin in the hybridization between the Fe $3d$ and the As $4p$ states.

\begin{figure} [htbp]
\includegraphics* [width=\columnwidth] {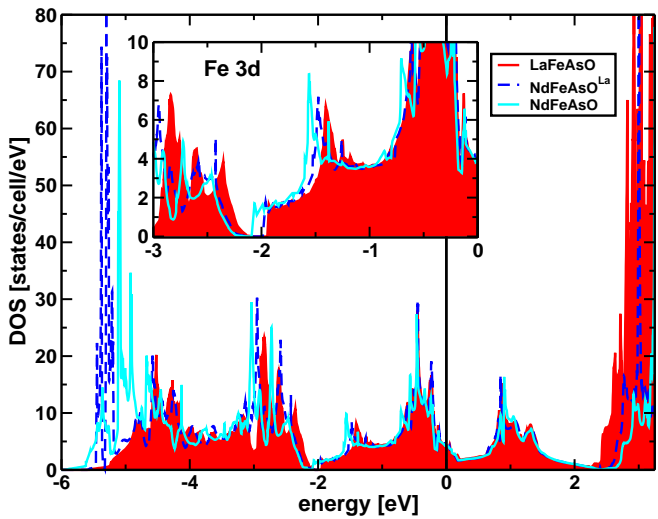}
\caption{(Color online) Total density of states for LaFeAsO (shaded red) and NdFeAsO (full light blue line) using the experimental room-temperature structural data. The density of states of the fictitious NdFeAsO) (labeled NdFeAsO$^{\rm La}$, using set~1 in Table.~I) is shown  by the dashed dark blue line. Inset: the density of states of the Fe $3d$ states for the same three structures.}
\label{FigureDOS}
\end{figure}

This result also nicely corroborates with the band-structure calculations and the density of states (DOS). In Fig.~\ref{FigureDOS} the total DOS for LaFeAsO and NdFeAsO (using the room-temperature experimental structural parameters) and the  fictitious Nd compound (labeled NdFeAsO$^{\rm La}$ and calculated by using the experimental structure of the La compound at room temperature) is shown.  The main features of the total DOS are very similar for the La and the fictitious Nd$^{\rm La}$ compounds---except for the occupied Nd $4f$ band (around $-5$~eV). Shorter lattice parameters in the Nd compound (going from the dashed to the full line in Fig.~\ref{FigureDOS}) have mainly an effect on the position of the $4f$ and the Fe $3d$ states (see below). Like in the La compound, we find also in the Nd compound a strong hybridization of Fe $3d$ with As $4p$ states in the energy region between $-4$ and $-2$~eV. The differences in these DOS can be seen more clearly on a smaller scale. The Fe $3d$ states are slightly shifted in direction of lower energy from the La via the fictitious Nd$^{\rm La}$ to the Nd compound; see inset of Fig.~\ref{FigureDOS}. These subtle electronic structure changes are also observed in the $^{75}$As NMR experiment. We notice that the As-Fe isotropic hyperfine coupling constant $a_{\rm FeAs}^{iso}=27.7\,{\rm to}\,29~{\rm kOe}/\mu_B$ is about 10\% larger than that reported for the La-family. This constant depends on the degree of polarization of the As $4s$ electrons caused by the As $4p$ states via exchange polarization effect. \cite{Tou} Therefore it may be---similar to the DOS---sensitive to the small crystalline structure variations such as lattice contraction or the shift of the As along the $z$ axis found for NdFeAsO$_{1-x}$F$_x$ in our x-ray diffraction (XRD) experiments. We conclude that all $^{75}$As NMR parameters, $a_{xz}$, $a_{\rm FeAs}^{\rm iso}$ and $\nu_Q$ are very sensitive to the details of As $4p$ electron distribution. Related to that, we refer here to the work of Mukuda et al. \cite{HM} where the monotonic increase in the superconducting transition temperature $T_C$ with $^{75}$As quadrupole frequency $\nu_Q$ has been reported. Based on these observations, our calculations (Table~I) may suggest that it is the lattice contraction and not the presence of additional RE $4f$ states that shifts $T_C$ to higher values in NdFeAsO$_{1-x}$F$_x$ compounds. This is also in line with pressure experiments on LaFeAsO, where $T_c$ was increased up to 43~K.\cite{tri}

Finally we comment on the  non-negligible coupling between $^{75}$As and Nd $4f$ moments, $a_{\rm Nd}^{\rm iso}=3.8\, {\rm kOe}/\mu_B$, which directly evidence an  interaction between the REO and FeAs layers. Such interlayer coupling could open the possibility for an indirect exchange interaction between the localized Nd $4f$ moments mediated by the Fe $3d$ and/or Nd $5d$ (partially occupied in the band-structure calculations) conducting electrons, i.e., a Ruderman-Kittel-Kasuya-Yosida (RKKY)-type interaction. The small Curie-Weiss temperature ($\theta=-14~{\rm K}$) as well as the low ordering temperature for the Nd moments may be due to the long-range nature of the RKKY-type interaction and its oscillatory dependence on the distance between the Nd moments. Additional experiments are needed to shed more
light on the size of Nd ordered moment at low temperatures.

\section{summary}

In summary, we have performed a study of the structural, electronic and magnetic properties of the superconducting NdFeAsO$_{1-x}$F$_{x}$ ($T_C=43\, {\rm K}$ for $x=0.15$) by high-resolution synchrotron x-ray powder diffraction, magnetization, and $^{75}$As NMR measurements. Structural investigations showed a contraction of the lattice structure with shorter RE-As bond distances, an increased thickness of the As-Fe$_2$-As layer and less distorted Fe-As$_4$ tetrahedra for the NdFeAsO family when compared to the LaFeAsO family. A typical $^{75}$As quadrupole powder line shape measured for NdFeAsO yielded an enhanced quadrupole frequency with respect to the La analogs. Furthermore, we calculated the electric field gradient using the band-structure code FPLO and obtained good agreement for the undoped compound. The enhanced EFG is due to a more prolate As $4p$ electron distribution, which is mainly caused by reduced lattice parameters and not the presence of Nd $4f$ electrons. We also found a scaling of the $^{75}$As NMR shift with the magnetic susceptibility of localized Nd moments suggesting a weak coupling between the NdO and FeAs layer. The RKKY- interaction between localized Nd $4f$ moments mediated by itinerant Fe $3d$ and/or Nd $5d$ states may originate from such a coupling.

\begin{acknowledgments}
J.W.G.B acknowledges the Royal Society of Edinburgh for financial support. The EPSRC-GB is thanked for provision of the beam time at the ESRF.
\end{acknowledgments}


\begin{thebibliography}{}
\bibitem{therm}
The temperature during the room-temperature scan was not controlled so it may have varied for a couple of degrees. Taking a typical value for a thermal expansion coefficient we note that the actual accuracy of the lattice parameters may be less than the ones obtained directly from the Rietveld refinement procedure.
\bibitem{supercell}
The supercell calculations were performed for the La compound (Ref.~\onlinecite{LaHajo}). Since the EFG is calculated on the As site and the substituted site is oxygen, a very similar behavior is expected for the Nd compound and there is no need for repeating the supercell calculation.
\end{thebibliography}
\end{document}